# Negating Negative Heat Capacity in Nanoclusters

Karo Michaelian[1] and Iván Santamaría-Holek[2]

[1] Instítuto de Física, Universidad Nacional Autónoma de México, Apdo. Post. 20-364, México D.F., 01000 México.

[2] Departamento de Física, Facultad de Ciencias, Universidad Nacional Autónoma de México, México D.F., 04510 México.

It is shown that "negative heat capacity" in nanoclusters is an artifact of applying equilibrium thermodynamic formalism on a "small" system trapped out of equilibrium in a particular structural motif representing only part of the energetically available phase space volume. Trapping may occur in either the canonical or microcanonical ensemble, but it is unavoidable in the microcanonical. A more general consequence of trapping is that all macroscopic quantities determined for nanoclusters will depend on the initial conditions.

36.40.Ei, 36.40.Mr, 36.40.-c

A system having negative heat capacity would demonstrate peculiar behavior. For example, it would cool down as energy were added or, equivalently, the system would heat up if energy were removed. Such behavior is in contradiction to experience and, in fact, violates one of the main symmetry principles of equilibrium thermodynamics; that of thermal stability. Nevertheless, negative heat capacity has been postulated, and even reportedly measured, in systems such as gravitating gases (stars and star clusters )[1], in the fragmentation of nuclei [2], atomic nanoclusters [3,4,5,6,7] and magnetically self-confined plasma [8].

All the systems, for which negative heat capacity has been reported, although ranging in size from the microscopic to the astronomical, are *small* systems in the sense that the range of the relevant interaction is greater than the characteristic size of the system. A feature of small systems is the presence of large energy barriers separating different regions of the energetically available phase space. As a consequence, for sufficiently low energies, the system becomes trapped and does not readily (or may never) attain thermodynamic equilibrium. In this letter we demonstrate that negative heat capacity in nanoclusters is an artifact of applying equilibrium thermodynamic formalism to a cluster trapped within a particular structural motif which is separated from other energetically available motifs by large energy barriers reaching into the solid to liquid transition region.

Although negative heat capacity in clusters has been critically examined before [9,10,11], there still lacks a general understanding of this anomaly, particularly in relation to energy barriers, trapping, and initial conditions. It is relevant that there are now a growing number of publications reporting negative heat capacity where the inconsistency is often presented as an inherent thermodynamic property of the cluster and characterized as a surprising[4] and even bizarre [12].

The second law of thermodynamics establishes that an isolated system (a member of a *microcanonical* ensemble) will naturally evolve towards a global maximum of the entropy function $S$ which is independent of the initial conditions, or, in other words, to the macropartition corresponding to the greatest multiplicity of the microstates. It follows that in a thermodynamic equilibrium state, $(\delta S)_{eq} = 0$ and $(\delta^2 S)_{eq} < 0$ where $\delta$ denotes an infinitesimal variation. Starting from Gibbs' equation, it is possible to show [13] that at constant volume $V$ and particle number $N$,

$$\delta^2 S = -\left(\frac{\partial E}{\partial T}\right)_{V,N} \frac{(\delta T)^2}{T^2} < 0 \quad \text{giving that } C_{V,N} \equiv \left(\frac{\partial E}{\partial T}\right)_{V,N} > 0 \;. \tag{1}$$

The heat capacity is thus positive definite, conferring thermal stability to the equilibrium state.

In the *canonical* ensemble, the system is in contact with a heat bath in equilibrium at constant temperature and the heat capacity at constant volume and particle number can be obtained from the second moment of the energy fluctuations[14] as $C_{V,N} = \frac{1}{k_B N T^2} \left\langle \left(E - \langle E \rangle\right)^2 \right\rangle$ which is also positive definite. It is of course reassuring that this equilibrium thermodynamic result does not depend on the ensemble used to facilitate the evaluation of the macroscopic properties. Similar conclusions can be obtained for the heat capacity at constant pressure $P$ and particle number $N$, more amenable to experimental measurement, since $C_{P,N} > C_{V,N} > 0$ [14].

Consider now a system in which the presence of an energy barrier divides its total phase space volume $\Gamma$ into two regions (see figure 1).

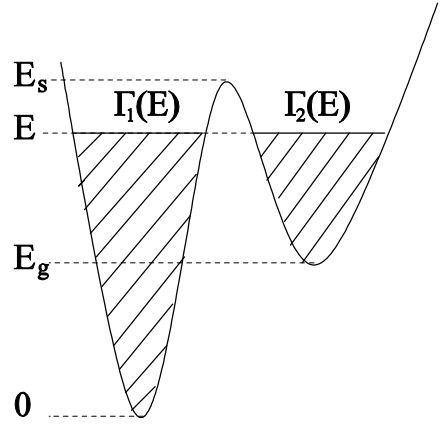

**Fig. 1: Schematic representation of the volume of phase space of a small system divided into two regions by a potential energy barrier.**

As shown in figure 1, $\Gamma(E)$ will in general depend on the total energy $E$ of the system in the form

$$\Gamma(E) = \Gamma_1(E) + \Gamma_2(E), \qquad (2)$$

where $\Gamma_1$ and $\Gamma_2$ are the volumes of the two subspaces separated by the energy barrier. Once given the phase space volume (2), the macroscopic properties of the system such as the microcanonical temperature $T$ and the heat capacity $C$ at constant parameters (for example pressure and particle number, assumed implicitly constant in the following) can be derived from the Boltzmann definition of the entropy

$$S = k_B \ln \Gamma. \qquad (3)$$

The microcanonical temperature $T$ is defined as

$$T^{-1} \equiv \partial S/\partial E = k_B \Gamma^{-1} \partial \Gamma/\partial E, \qquad (4)$$

which, in view of (2), leads to

$$T = \frac{\Gamma_1 + \Gamma_2}{k_B} \left\{ \frac{\partial \Gamma_1}{\partial E} + \frac{\partial \Gamma_2}{\partial E} \right\}^{-1}. \qquad (5)$$

The microcanonical temperature (5) is consistent with the thermodynamic formalism and, for a system in equilibrium, has exactly the same relation with energy as the temperature determined within the canonical ensemble.

The central problem is that $\Gamma(E)$ is not *a priori* known but must be determined either experimentally or through simulation. In the case of small systems such as nanoclusters, a difficulty arises because the presence of the energy barrier between regions **1** and **2** of the phase space hides the existence of region **2** to the experimenter or simulator confined to region **1** (or vice versa) until the energy barrier has been surmounted. Then, instead of obtaining a phase space volume as given by (2), an experimenter or simulator working in the microcanonical ensemble will determine a discontinuous phase space volume

$$\gamma_i(E) = \Gamma_i + \Gamma_j \Theta(E - E_s), \qquad (6)$$

where $i,j = \mathbf{1,2}$ ($j \neq i$) depending on the initial conditions, $\Theta(E - E_s)$ is the Heaviside function, and $E_s$ is the height of the energy barrier (see Fig. 1). Note that because of the factor $\Theta(E - E_s)$, Eq. (6) violates the equiprobability assumption of equilibrium statistical mechanics, fundamental for establishing the connection with equilibrium thermodynamics. However, using anyway Eq. (4) and (6) would lead to a corresponding determined "temperature" of

$$t_i = \frac{\left(\Gamma_i + \Gamma_j \Theta(E - E_s)\right)}{k_B} \left\{ \frac{\partial \Gamma_i}{\partial E} + \frac{\partial \Gamma_j}{\partial E} \Theta(E - E_s) + \Gamma_j \delta(E - E_s) \right\}^{-1}. \qquad (7)$$

The salient points of this result are;

1. Because of the Heavyside function appearing in the denominator, $t_i$ is not necessarily a monotonically increasing function of the energy.

2. For energies greater than $E_s$ or less than $E_g$, where $E_g$ is the energy gap between regions **1** and **2** (see Fig. 1), the determined "temperatures" $t_i$ are equal to the microcanonical temperature (5) and independent of the initial conditions.

3. At the energy $E = E_s$ the phase space volume changes discontinuously and, as a consequence, there is a discontinuity and an in-determination in the determined "temperature" as represented by the Heavyside and Dirac delta functions respectively in (7).

4. For energies lying in the range $E_g < E < E_s$ the determined "temperatures" $t_i$ are different from the microcanonical temperature $T$ and depend on the initial conditions. For example, starting in region **1,** and assuming that $\Gamma_1 \gg \Gamma_2$ (which will be the case for any non-negligible $E_g$) would lead to a determined temperature $t_1$ of

$$t_1 \cong T\left(1 + \frac{\Gamma_2'}{\Gamma_1'}\right) \qquad \text{for } E_g < E < E_S, \tag{8}$$

where $\Gamma_{i,j}' \equiv \partial \Gamma_{i,j}/\partial E$. The determined temperature $t_1$ will thus be greater than the microcanonical temperature $T$. Therefore, since $t_1 > T$ for $E_g < E < E_s$ and $t_1 = T$ for $E > E_s$, with $T$ a continuously increasing function of the energy, it follows that $t_1$ must decrease after the barrier at $E = E_s$ is surmounted. This means that, *if* we take the variable $t_1$ as the thermodynamic temperature, on surmounting every lowest energy barrier connecting two minima of a nanocluster we should observe "negative heat capacity". However, this effect will normally not be observable because of the differences in the contribution of the kinetic density of states due to the energy gap $E_g$. For example, taking $\Gamma_1 \propto E^\alpha$ and $\Gamma_2 \propto (E - E_g)^\alpha$ gives $\Gamma_2'/\Gamma_1' \propto (1 - E_g/E)^{\alpha-1}$ which, for any nonzero $E_g$, is normally very small since $\alpha$ is of order $3N$ where $N$ is the number of atoms in the cluster. However, the density of states is composed of both a kinetic and a configurational contribution, and if the configurational contribution to $\Gamma_2'$ could compensate the larger kinetic contribution to $\Gamma_1'$, then $t_1$ may be significantly larger than $T$ and its decrease to $T$ may become observable on passing the barrier at $E_s$. Indeed, we will see that "negative heat capacity" may be determined on surmounting the particularly high energy barriers separating isomers of one structural motif (symmetry) from those of another if these barriers reach into the transition region where the configurational density of states becomes suddenly very large.

Likewise, for the determined temperature starting in region **2**, we find (for $\Gamma_1 \gg \Gamma_2$),

$$t_2 \cong T\left(\frac{\Gamma_1'}{\Gamma_1} \frac{\Gamma_2}{\Gamma_2'}\right).$$ Using the example above of the power law dependence of the density of states

on the energy gives

$$t_2 \cong T\left(1 - \frac{E_g}{E}\right) \qquad \text{for } E_g < E < E_S. \qquad (9)$$

$t_2$ will therefore be less than the microcanonical temperature $T$.

On the basis of the results obtained above for our model phase space volume, we can now explain the "caloric curves" determined in the microcanonical ensemble for LJ$_7$ (Fig. 2(a)), and Na$_{147}$ (see Ref. [15]).

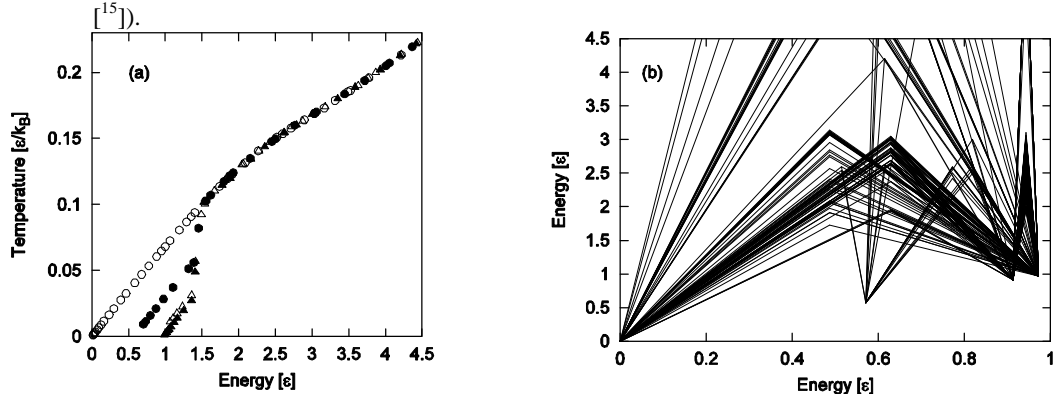

**Fig. 2: (a) "Caloric curves" for LJ$_7$ (7 atom Lennard Jones cluster), obtained using constant energy molecular dynamics (different symbols correspond to initial conditions in different minima). (b) The most important lowest energy saddles connecting the 4 minima of LJ$_7$ obtained through a statistical algorithm [16]. The x-axis of the saddles is arbitrarily set to the average value of the energy of all minima they connect.**

From Fig. 2(b), which plots the most important saddles connecting minima of LJ$_7$, it can be seen that below an energy of $1.5\varepsilon$, the barriers would prevent the system from ever reaching thermodynamic equilibrium at constant energy. Therefore, in accordance with incise 4 above, for $E_g < E < E_s \approx 1.5\varepsilon$, the determined "temperature" $t_1$ will be greater than the thermodynamic temperature (although imperceptibly so since $\Gamma_2'/\Gamma_1' \ll 1$, while $t_2, t_3, t_4$ (corresponding to initial conditions starting in the other isomers) will be less than the thermodynamic temperature and given by Eq. (9) with the energy gaps from the global minimum (the $E_g$'s) taken from figure 2(b). There will be an in-determination in the "temperatures" at $E = E_s \approx 1.5\varepsilon$ due to the discontinuous increase in $\Gamma$, and above this energy all minima become accessible and all determined "temperatures" will then be equivalent to the thermodynamic temperature. This is exactly what constant energy molecular dynamics gives using the 4 distinct minima as initial conditions (Fig. **2**(a)), demonstrating that molecular dynamics at constant energy gives the initial condition dependent determined "temperatures" $t_i$ and not the thermodynamic temperature $T$.

Figure 3(a) shows the energy distribution of the lowest energy isomers and the interconnecting saddles for $Na_{147}$. There are two principle structural motifs for the minima, the icosahedral Ih (in dark grey, red on-line) and the decahedral Dh (in grey, green on-line). The lowest energy inter-motif barriers are of the oder of 1.5 to 2.5 eV while the lowest barriers found between different structural motifs are of the order of 7.8 eV (in black, blue on-line). Using this data, we can construct the total density of states using the superposition principle[17][18], and thereby the caloric curve, as obtained by being both insensitive and sensitive to the barriers via a straight forward many-minima generalization of Eqs. (5) and (7) respectively.

The caloric curve insensitive to the barriers, Fig. 3(b) (dashed line), shows a smooth continuous transition region with positive slope. The caloric curve sensitive to the barriers, Fig. 3(b) (solid line), clearly shows that a negative heat capacity is determined at the height of the energy barrier separating the icosahedral from the decahedral structural motifs, $E_s \approx 7.8$ eV. Also, the high barrier reduces the width of the transition region and alters (in this case raises) the transition temperature. The inter-motif barriers do not significantly affect the caloric curve, increasing only slightly the anomaly at the transition. This caloric curve is in good agreement with that obtained from constant energy molecular dynamic simulations employing the same Gupta potential, giving "negative heat capacity" at the same energy (see [15]). We thus conclude that the molecular dynamic runs are trapped in the icosahedral minima and are blind to the decahedral minima until reaching an energy of approximately 7.8 eV, even though the first decahedral isomer begins to contribute to the true phase space volume at only 0.84 eV (see Fig. 3(a)). Starting in the decahedral minima leads to the dotted curve in Fig. 3(b), demonstrating that below 7.8 eV, all results, including "negative heat capacity", depend on the initial conditions.

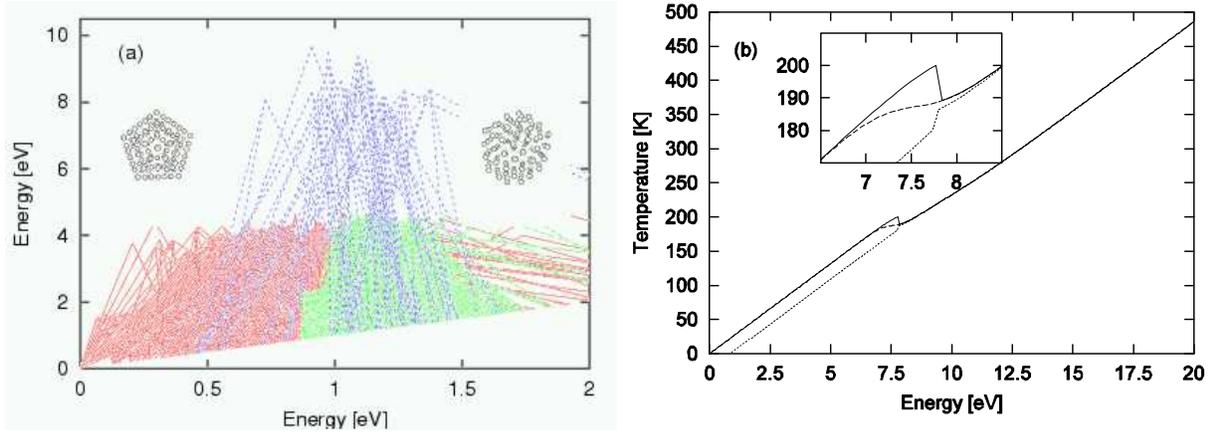

Fig. 3: (a) Lowest energy saddle-minima connections between icosahedral Ih (in dark grey, red on-line) and decahedral Dh (in gray, green on-line) minima for $Na_{147}$ (Gupta potential). The lowest Ih-Dh saddles found are also given (in black, blue on-line). (b) The caloric curves obtained by being both insensitive (dashed line) and sensitive to the barriers (solid line - initial conditions in the Ih minima; dotted line - initial conditions starting in the Dh minima) as obtained from a many-minima generalization [18] of Eqs. (5) and (7) respectively.

The most general condition for negative heat capacity, obtained from Eqns. (1) and (4), is $\Gamma''/\Gamma' > \Gamma'/\Gamma$ where $\Gamma'' \equiv \partial^2 \Gamma / \partial E^2$. (This gives rise to an inverted curvature, or negative intruder, in the entropy function.) This condition can only be satisfied for a continuous function $\Gamma(E)$ which increases faster than exponentially with energy. For a system of finite particle number in equilibrium, this is not possible since the phase space volume grows at most as a power law in energy, which is always less than exponential. Therefore, a discontinuous increase in the measured phase space volume is a necessary condition for determining "negative heat capacity". However, finding such a discontinuity already implies a lack of ergodicity. As we have shown, "negative heat capacity" will only become detectable if the discontinuous increase in the phase space volume is large, which will occur if the barriers reach into the phase transition where the phase space volume increases by orders of magnitude. Barriers lower or higher than the transition region will not give a detectable "negative heat capacity" but the melting transition may be incorrectly determined to be at the height of the barrier in the latter case.

It is now incumbent to discuss why "negative heat capacity" has been found in numerous other theoretical and experimental studies. The analytical model of Bixon and Jortner [19] assumes discrete energy states, implying discontinuous increases in the volume of phase space as a function of energy. In fact, their microcanonical density of states contains a Heavyside function, similar to that in our Eq. (6) resulting from the energy barrier. The determination of "negative heat capacity" for discontinuities large enough in the phase space volume then follows.

Labastie and Whetten [3], using a Monte Carlo technique, obtained a "back bending" in the microcanonical caloric curves for both a 55 and 147 atom Lennard Jones cluster after an inverse Laplace transformation from the canonical ensemble. Schmidt et al. [4] report an experimental determination of negative heat capacity for $Na_{147}$ after an inferred inverse Laplace transformation of their result to the microcanonical. Jellinek and Goldberg[5], using constant energy molecular dynamics with the Gupta potential, found a "back bending" in the caloric curve for $Al_{147}$, while we found a similar result for $Na_{147}$ [6]

In order to understand these findings in light of the results of this letter, we studied the potential energy landscapes of both the Lennard Jones and Gupta potentials using the statistical genetic algorithm described in reference [16]. For $Na_{147}$, as shown in Fig. 3(a), the barriers between the Ih and Dh motifs are at a height of 7.8 eV, in the transition region, while the decahedral isomers begin to contribute to the true thermodynamics at only 0.84 eV (Fig. 3(a)). Similarly, for $LJ_{55}$, barriers between the Ih and fcc (and between Dh and fcc) isomers were found at the transition region while the fcc isomers contribute to the true phase space volume at much lower energy. Therefore, at fixed energy below these high barriers, the sampling of phase space is necessasarily incomplete, implying a large discontinuity in the measured phase space volume on surmounting the barrier, and thus leading to the incorrect determination of "negative heat capacity" at the transition (see the caloric curves in references [3] and [15]).

With regard to theory and experiment performed in the canonical, in agreement with our analysis of Na$_{147}$, experiments on gold nanoparticles by Koga et al.[20] demonstrate extremely high energy barriers (close to the melting point) for the Ih to Dh transition while those of the Dh to fcc are even above the melting transition. Attaining ergodicity with these high barriers requires simulation times, or experimental ensemble sizes and thermalization times, many orders of magnitude larger than presently considered[21]. Further evidence for these systems being trapped out of equilibrium is the fact that in both theory and experiment for which "negative heat capacity" has been determined, the microcanonical caloric curves, as obtained from the inverse Laplace transformation of the canonical density of states is different from the canonical caloric curve, implying a sampling statistics in the canonical ensemble which is not Boltzmann. Finally, we note that a negative heat capacity for both LJ$_{13}$ and LJ$_{38}$ was initially suspected until more ergodic canonical methods were employed (see references [22,23]).

In conclusion, trapping is an *in principle* problem for nanoclusters analyzed in the microcanonical ensemble. Extremely high barriers exist between different structural motifs for nanoclusters and these barriers fragment the available phase space volume, leading to discontinuous increases in the measured density of states with energy. If these barriers reach into the transition region where the configurational density of states increases rapidly, "negative heat capacity", a reduced transition width, and an incorrect "melting temperature" may be determined on blindly applying equilibrium thermodynamics. The canonical ensemble is not exempt from these problems since these extremely high barriers imply that achieving thermodynamic equilibrium may be non-trivial. For trapped systems in general, all macroscopic results will depend on the initial conditions and will therefore have no thermodynamic significance. In light of these findings, a careful reanalysis of previous macroscopic results on nanoclusters may be in order.

*Acknowledgements*: The authors thank J.A Reyes-Nava and A. Taméz for supplying Fig. **2**(a). Financial support of project DGAPA IN104402 and IN108006 is gratefully acknowledged.